\documentclass[submission,Phys]{SciPost}

\usepackage[english]{babel}
\usepackage{graphicx}
\usepackage{dcolumn}
\usepackage{bm}
\usepackage{units}
\usepackage{amsmath}
\usepackage{amssymb}

\usepackage{color}
\usepackage[normalem]{ulem}
\usepackage{color}

\definecolor{bleufoncePython}{rgb}{0.2, 0.2, 0.6}
\definecolor{violetPython}{rgb}{0.46, 0, 0.63}
\definecolor{vertPython}{rgb}{0.04, 0.46, 0.09}
\definecolor{rougePython}{rgb}{0.55, 0.035, 0}
\definecolor{jaunePython}{rgb}{0.72, 0.52, 0.04}
\definecolor{bleuPython}{rgb}{0.0, 0.11, 0.5}
\definecolor{turquoisePython}{rgb}{0.0, 0.39, 0.45}

\begin{document}


\begin{center}{\Large \textbf{Asymptotic temperature of a lossy condensate }}\end{center}
\begin{center}
  I. Bouchoule\textsuperscript{*(1,2)}, 
  M. Schemmer\textsuperscript{(1,2)} and C. No\^us\textsuperscript{(2)}
\end{center}

 \begin{center}
(1)Laboratoire Charles Fabry, Institut d'Optique, CNRS, Université Paris-Saclay,
   2 Avenue Augustin Fresnel, 91127 Palaiseau Cedex, France\\
(2) Laboratoire Cogitamus, service publique de recherce et d'enseignement supérieur   
\\ 
* isabelle.bouchoule@institutoptique.fr
 \end{center}
 
\section*{Abstract}
{\bf We monitor  the time evolution of the temperature 
of phononic collective modes in a one-dimensional quasicondensate  
submitted to  losses. 
At long times the  ratio between the temperature and 
the energy scale $mc^2$, where $m$ is the atomic mass and $c$ the sound 
velocity takes, within a precision of 20\%, an asymptotic value.
This asymptotic value is observed while $mc^2$ decreases in time by 
a factor as large as 2.5. 
Moreover this ratio is shown to be independent on the 
loss rate and on the strength of interactions.
These results confirm theoretical predictions and 
the measured stationary ratio is in quantitative agreement 
with the theoretical calculations.
}

\section{Introduction}

There has been many effort and progress in the last decades for the realization and investigation
of isolated many-body quantum systems.
The effect of coupling to an environment has however regained
interest in the last years. While such a coupling was manly considered as detrimental for
the study of many-body quantum physics, it has been shown that proper engineering of the
coupling to an environment could enable the
realization of interesting many-body quantum states such as
entangled states or highly correlated states~\cite{barreiro_open-system_2011,tomita_observation_2017}.
The effect of coupling to an environment is still a widely open question.
The simplest kind of coupling, which is ubiquitous in experiments, is a loss process where the
particles leave the system. 
Losses are particularly relevant in exciton-polariton condensates but they are also present, or
can be engineered, in ultra-cold atomic degenerate Bose gases. 
If one considers 
a Bose-Einstein condensate (BEC) wavefunction,
losses are treated as a dissipative term added to the Gross-Pitaevskii
equation, equation which describes the evolution of the BEC at the
mean-field level.
This approach was
successful in describing the effect of local losses
in an atomic BEC~\cite{barontini_controlling_2013}
and many aspects of exciton-polariton
condensates~\cite{
  wouters_excitations_2007,wouters_spatial_2008,keeling_spontaneous_2008,lagoudakis_quantized_2008}.
In the last case, a pumping process ensures
the presence of a steady state. 
Beyond this mean-field approach, the loss process introduces fluctuations,
which are due to the
shot noise associated to the quantization of the particles. 
Both the dissipation and the fluctuations produced by losses was
taken into account in
stochastic theoretical descriptions~\cite{carusotto_spontaneous_2005,wouters_stochastic_2009,grisins_degenerate_2016,bouchoule_cooling_2018}\footnote{Other approaches such as the Keldish formalism
  have been developped~\cite{szymanska_nonequilibrium_2006}}

While for exciton-polariton condensates a pumping process is present,
in atomic Bose gases
the sole effect of losses can be investigated.
In~\cite{grisins_degenerate_2016,bouchoule_cooling_2018}
the time evolution of
a Bose-Einstein condensate, or a quasicondensate in 
reduced dimension, submitted to homogeneous losses has been
theoretically investigated.
 The dissipative term is responsible for  cooling:
  although the loss process is homogeneous,
  losses per unit length occur at a higher rate in regions of higher densities
  -- just because there are more atoms-- which leads to a decrease
  of density fluctuations and thus of their associated interaction energy.
  On the other hand, the
  stochastic nature of losses tends to increase density fluctuations and thus
  the interaction energy; this corresponds to a heating term. 
 As a result of the competition between  both effects, 
it has been predicted that phononic collective 
modes acquire, at large times, a temperature $k_BT$ that decreases 
proportionally to the energy scale $mc^2$ where $m$ is the 
atomic mass and $c$ the speed of sound.

The precise value of the 
asymptotic ratio $k_B T/(mc^2)$ depends on the loss process
and the geometry~\cite{bouchoule_cooling_2018}. An intrinsic
homogeneous loss process present in 
cold atoms setup is a three-body loss process where a loss event
corresponds to an inelastic collision involving three atoms and amounts 
to the loss of the three atoms. In~\cite{schemmer_cooling_2018},
the asymptotic ratio  $k_B T/(mc^2)$
associated to this three-body losses has been experimentally observed,
and its value
is  in agreement with theoretical predictions. 
On the other hand, there was up to now no experimental evidence
of an asymptotic ratio $k_B T/(mc^2)$
in the case of a one-body loss process\cite{rauer_cooling_2016}. A
one-body loss process corresponds  to
a uniform loss rate: each atom has the probability $\Gamma dt$ to be lost
during a time-interval $dt$,
regardless both of its position and its energy.
In this paper, we demonstrate the presence of an
asymptotic value of the ratio $k_BT/(mc^2)$ for one-dimensional
harmonically confined
quasicondensates submitted to one-body losses, and our results are in agreement
with theoretical predictions.

\section{Description of the experiment and data analysis}
We use an atom-chip set-up, described in detail
in~\cite{schemmer_out--equilibrium_2019}, to produce
ultracold gases of $^{87}$Rb atoms, polarized in the stretch state $|F=2,m_F=2\rangle$ and
confined in a very elongated magnetic trap.
The transverse confinement is realized  by three parallel wires aligned along $z$,
running an AC current modulated at 400~kHz, together with a homogeneous longitudinal magnetic
field $B_0=\unit[2.4]{G}$~\cite{trebbia_roughness_2007}:
atoms are confined transversely in the time-averaged potential and the transverse
oscillation frequency $\omega_\perp/(2\pi)$,
which depends on the data-set, lies in the interval [1.5-4.0]$\,$kHz. 
A longitudinal harmonic confinement of frequency $\omega_z/(2\pi)=\unit[9.5]{Hz}$ is realized by a pair of wires
perpendicular to $z$.
Using standard radio-frequency (RF) evaporative cooling,
we prepare  clouds whose temperature $T$ and chemical
potential\footnote{The energy offset used for chemical potential is the energy of the
transverse ground state, {\it i.e.} $\hbar\omega_\perp$.} $\mu_p$, depending on the data set, lie in the range $\mu_p/(2\pi\hbar)\in \unit[1.0-3.1]{kHz}$
and $T  \in \unit[40-75]{nK}$.
The ratios $k_B T/(\hbar \omega_\perp)$ and $\mu_p/(\hbar\omega_\perp)$
lie in the range $0.3-0.7$ and $0.6-1.2$ respectively such that the clouds
are quasi-one-dimensional.
The clouds lie deep in the quasicondensate regime~\cite{kheruntsyan_pair_2003},
which is characterized
by strongly reduced density fluctuations --the
two-body zero distance correlation function $g^{(2)}(0)$ being
close to one -- while  longitudinal phase fluctuations are still present. 
We then increase the frequency of the RF knife by about 25 kHz, so that it no longer induces losses
but ensures the removing of residues of three-body recombination events.

In contrast to the three-body process, no intrinsic process leads to
a  one-body loss term in our experiment and one-body losses have to be
engineered. We introduce homogeneous one-body losses by coupling the trapped atoms to
the untrapped state $|F=1,m_F=1\rangle$, which 
lies at an energy $\hbar\omega_{HFS}+(3/2)\mu_B B_0$
below the trapped state $|F=2,m_F=2\rangle$, where $\omega_{HFS}\simeq \unit[6.8347]GHz$
is the hyperfine splitting of the $^{87}$Rb ground state.
Coupling is realized by  a microwave (MW) field 
produced by a voltage-controlled oscillator connected to an antenna placed
a few centimeter away from the
atomic cloud. We use a noise-generator to
produce a MW power spectrum
which presents a rectangular shape 200~kHz wide.
Its central
frequency $\omega_0$ may be varied in time. During the preparation phase of our ultra-cold cloud,
$\omega_0$ is chosen such that the transition is shifted from resonance by about 5~MHz so that the
MW does not induce any noticeable losses. At time $t=0$, we suddenly shift $\omega_0$ to its
resonance value 
to induce losses.
The large width of the MW power spectrum,
compared to $\omega_\perp$ and to the chemical potential of the atoms,
ensures that the loss rate is  homogeneous over the size of the atomic cloud and is
not affected by interaction effects. 
We adapt the loss rate $\Gamma$ adjusting the power of the MW field.

\begin{figure}
  \centerline{\includegraphics[width=0.9\textwidth]{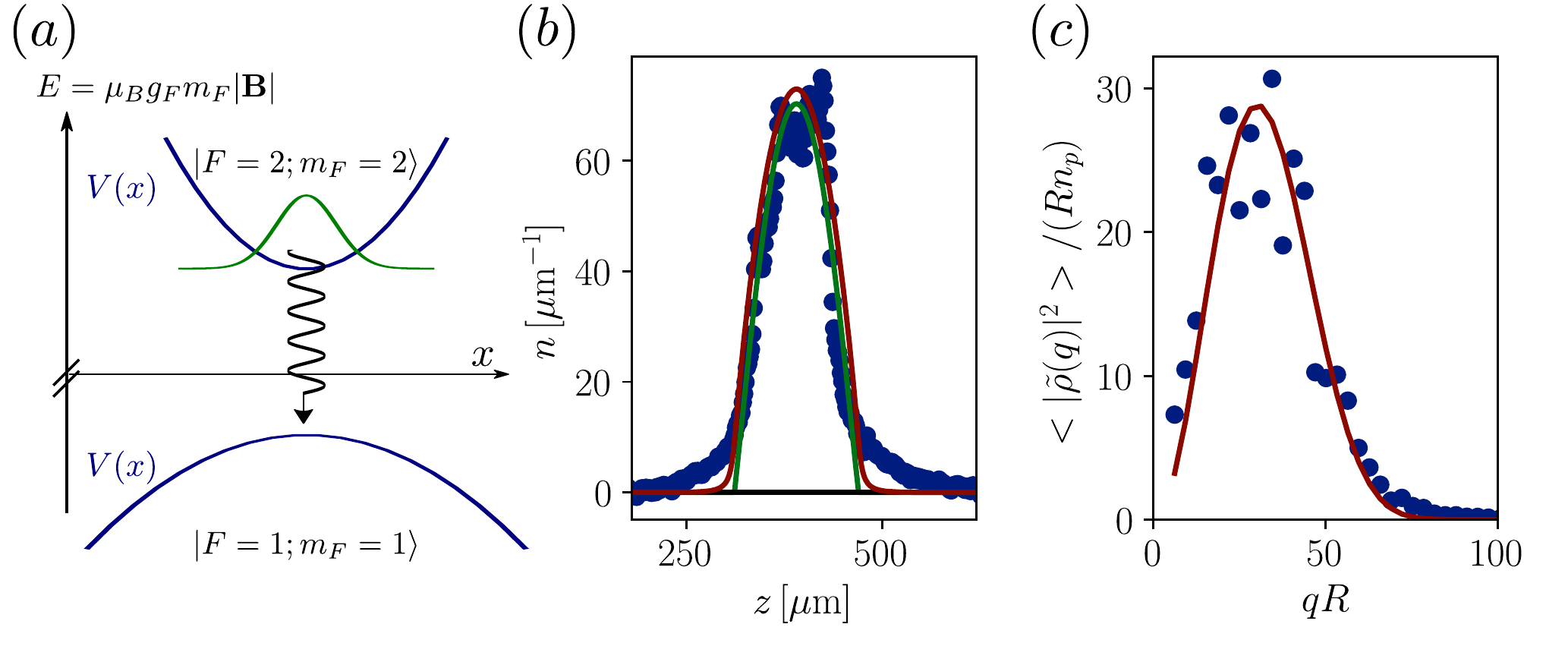}}
  \caption{Implementation of one-body losses in our magnetically-confined gas and data analysis.
  (a) sketch of the MW coupling between the trapped and untrapped state.
    (b) A typical density profile $n_0(z)$, obtained averaging
    over about 20 images. The
    data to which it corresponds (5$^{\rm{th}}$ data point of data set 6)
    is the encircled data point in Fig.(a).
    The shape expected for a
    quasicondensate is shown in green, with the peak density $n_p$
    as single fitting parameter.
  (c) Density ripples power spectrum for the same data as (b),
    together with the theoretical fit yielding the temperature $T$.
    The red solid line in (b) is the density profile expected for a cloud at a temperature $T$
  using the Yang-Yang equation of state and the local density approximation.}
  \label{fig.drawing_analysis}
  \end{figure}

We analyze the atomic cloud using absorption images taken after a time of flight
$t_f=\unit[8]{ms}$ following  the sudden switch off of the  confining
potential. We acquire an ensemble of about 20 images taken in the same experimental
conditions.
The fast transverse expansion of the cloud
provides an effective instantaneous switch off of the interactions with respect to the longitudinal
motion and the gas evolves as a non-interacting gas for $t_f$.
Averaging over the data set, 
we extract the longitudinal  density profile $n_0(z)$.
The longitudinal velocity width, of the order of $k_B T /(\hbar n_p)$, where $n_p$ is the
peak linear density~\cite{mora_extension_2003}, is small enough so that
the longitudinal density profile is not affected by the time-of-flight and $n_0(z)$ is equal
to the density profile of the cloud prior to the trap release.
From $n_0(z)$, we extract
the total atom number and the peak density $n_p$. 
The latter is obtained by fitting the central part of the measured density profile
with the profile expected for a gas lying in the  quasicondensate regime.
To compute the quasicondensate density profile we rely on the local density
approximation (LDA): the gas at position $z$ is described by
a homogeneous gas at chemical potential $\mu(z)=\mu_p -  m\omega_z^2z^2/2$,
and the linear density is derived from $\mu(z)$ using
the equation of state of a homogeneous quasicondensate. The latter, which
relies the chemical potential
  $\mu$ to the linear density $n$,  is $\mu=\hbar\omega_\perp(\sqrt{1+4n a_{3D}}-1)$,
where $a_{3D}$ is the 3D scattering length~\cite{salasnich_effective_2002,fuchs_hydrodynamic_2003}. 
For $na_{3D}\ll 1$ it  reduces to the pure 1D expression $\mu=g_{1D}n$ , where
 the 1D coupling constant\footnote{In our
   case, $\omega_\perp\ll \hbar/(ma_{3D}^2)$ such
   that we are far from confinement-induced resonance.} is 
 $g_{1D}=2\hbar\omega_\perp a_{3D}$~\cite{olshanii_atomic_1998}. At larger $n a_{3D}$ it includes
the effect due to the transverse swelling of the wavefunction.
The longitudinal quasicondensate profile extends over $2R$,
where $R= \sqrt{2\mu_p/(m\omega_z^2)}$.
Fig.~(\ref{fig.drawing_analysis})(b) shows a typical experimental density
profile $n_0$, together
with the theoretical quasicondensate  profile. 
The  good agreement between most of the cloud's shape and
 the calculated profile confirms that
the cloud lies deep into the quasicondensate regime. It also confirms
that the loss rate
is small enough so that the cloud shape has time to follow adiabatically
the atom number decrease.

Temperature determination is realized by the well-established
density-ripple thermometry method~\cite{dettmer_observation_2001,imambekov_density_2009,manz_two-point_2010,schemmer_monitoring_2018,schemmer_cooling_2018}.
This thermometry uses the fact that thermally
excited phase fluctuations initially present in the cloud transform into density fluctuations
during $t_f$ such that single shot images of  the cloud presents large random density ripples.
From the set of acquired images, we extract the power spectrum of the density ripples.
More precisely,
we extract from
each image  $\rho_q=\int_{-R}^{R} dz \delta n(z)e^{iqz}$ where $\delta n(z)=n(z)-n_0(z)$
and we then compute the density ripple power spectrum $\langle |\rho_q|^2\rangle$,
from which we remove
the expected
flat-noise contribution of optical shot noise.
The power spectrum is then fitted with the expected power spectrum for a quasicondensate
of peak density $n_p$ confined in a harmonic longitudinal potential, calculated using
the LDA approximation~\cite{schemmer_monitoring_2018}, with the
temperature as fitting parameter\footnote{ We take into
account the finite imaging resolution by multiplying the theoretical power spectrum with
$e^{-k^2\sigma^2}$ where $\sigma$ is the rms width of the imaging point-spread-function. 
Due to finite depth-of-focus, $\sigma$  depends on the size of the cloud along the imaging axis,
which itself depends on $\omega_\perp$. Thus $\sigma$ may depend on the data set but
for a given data set we use the same $\sigma$ for all evolution times. }.
Fig.~\ref{fig.drawing_analysis}(c) shows an example of a power spectrum (corresponding to the encircled
data point in Fig.~(\ref{fig.evolNat})(a)), together with the theoretical fit yielding
the temperature $T$.
 This thermometry probes fluctuations whose
wavelengths are much larger than the healing length $\xi=\hbar/\sqrt{m g n_p}$ such that
the temperature corresponds to the temperature of the phononic collective modes.

\section{Experimental results}
We investigate the time evolution of the atomic cloud for
6 different data sets which correspond to different
transverse oscillation frequencies -- {\it i.e.} different interaction 1D effective coupling constant --,
different initial situations -- {\it i.e.} different atom number and temperature -- and
different MW power -- {\it i.e.} different 1-body loss rate. They are listed in table~(\ref{table.data}).

\begin{table}
\centering
\begin{tabular}{|l|l|l|c|}
  \hline
      {\bf data-set number}&{\bf $\omega_\perp/(2\pi)$ (kHz)}& {\bf $\Gamma$ (s$^{-1}$)}&{\bf Symbol}\\
        \hline  \hline
  1 & 1.5 & 3.8 &{\color{jaunePython} $\blacktriangleleft$} \\ \hline
  2 & 1.5 & 1.6 & {\color{turquoisePython} $\bigstar$}\\ \hline
  3 & 2.1 & 5.2 & {\color{bleuPython} \raisebox{-0.08cm}{{\huge $\bullet$}}} \\  \hline
  4 & 3.1 & 4.9 &  {\color{vertPython} $\blacksquare$} \\  \hline
  5 & 3.1 & 2.5 &  {\color{rougePython} $\blacktriangle$} \\  \hline
  6 & 4.0 & 4.5 &{\color{violetPython} $\blacktriangleright$}\\  \hline
  
\end{tabular}  
\caption{Data sets presented in this paper, with the associated symbol used in the figures 2 and 3.}
\label{table.data}
  \end{table}

\begin{figure}
  \centerline{\includegraphics[height=6cm]{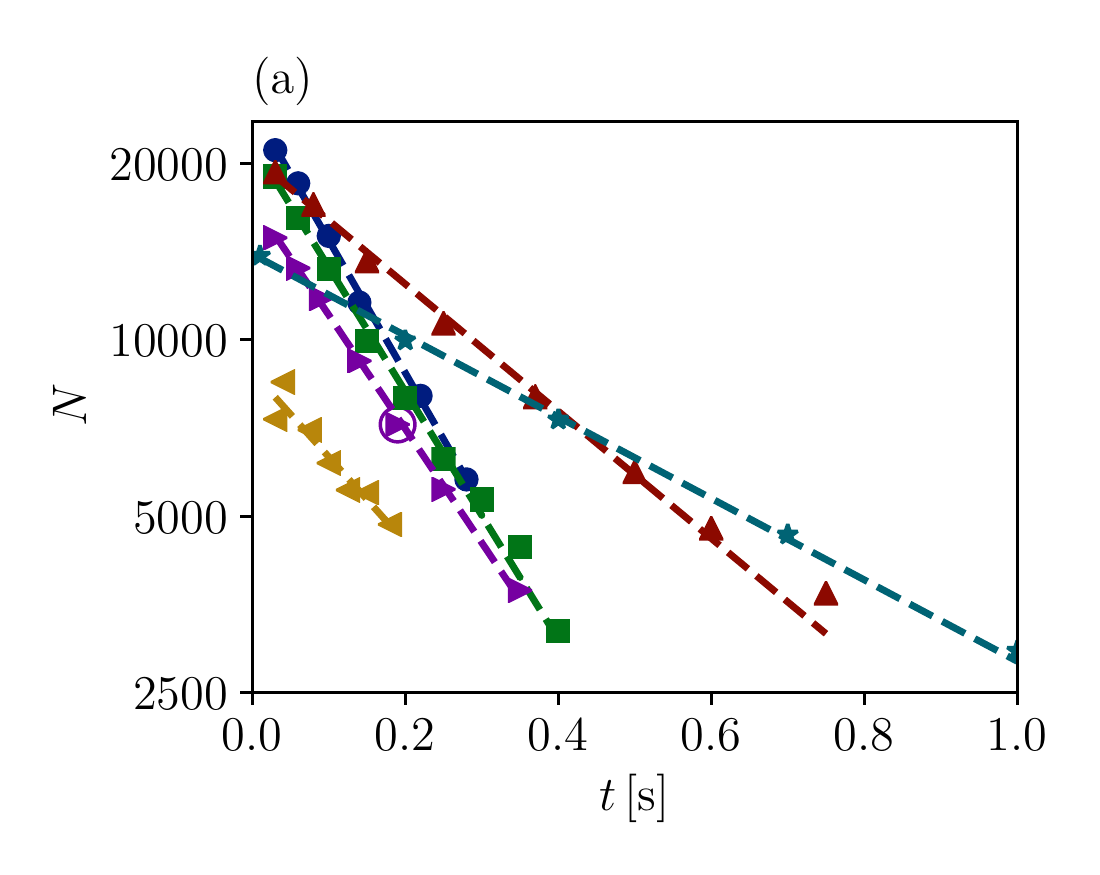}
    \includegraphics[height=6cm]{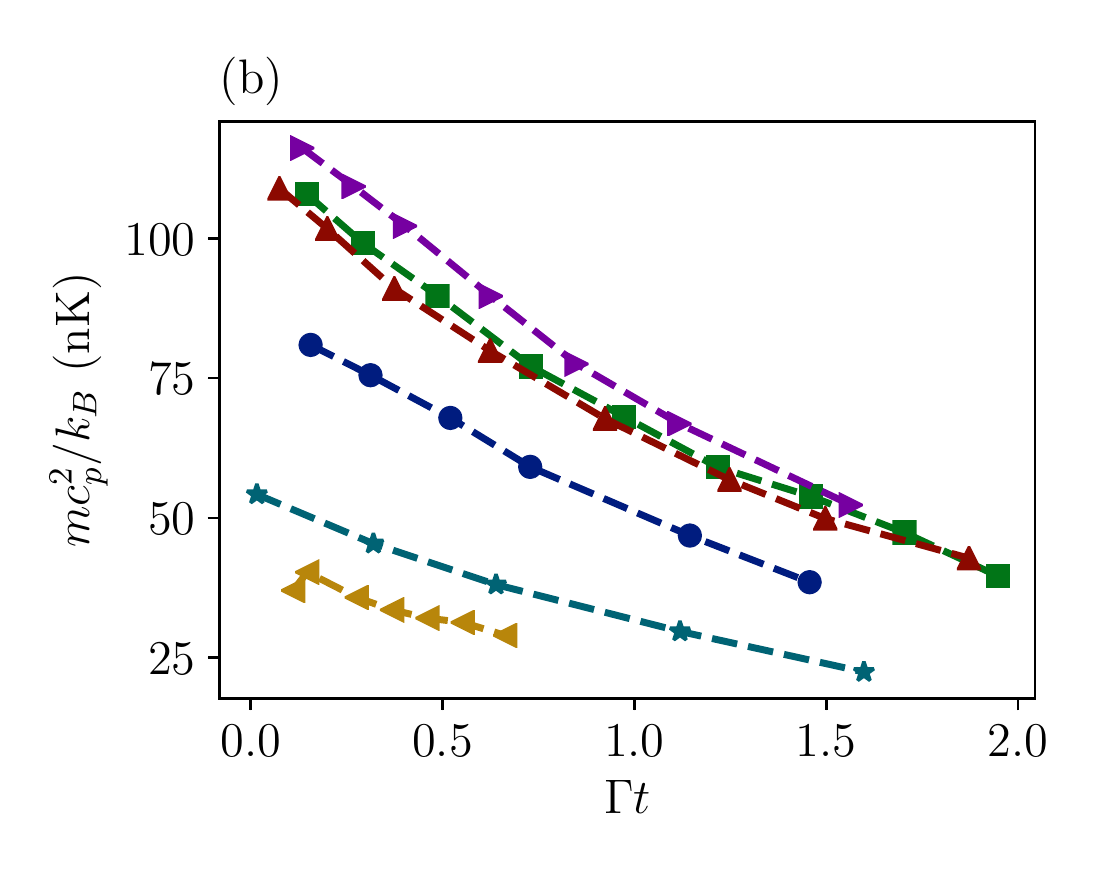} 
}
  \caption{(a) Evolution of the total atom number for the different sets of data shown in semi-log scale.
    The loss rate $\Gamma$ for each data set is deduced from an expontential fit, shown as dashed line. 
    (b) Evolution of the energy scale $mc_p^2$ for same data sets as a function of
    $\Gamma t$. Both plots use the color codes of table.~\ref{table.data}
    }
  \label{fig.evolNat}
\end{figure}

We plot in Fig.~(\ref{fig.evolNat})(a) the time evolution of
the total atom number for the different data
sets.
The exponential decrease of the atom number, shown by the  good agreement with 
exponential laws represented by strait lines in the semi-log plot,
confirms that we realized a uniform  one-body loss process. 
The loss rate varies by
roughly a factor 2 between different data sets. As pointed out in
the introduction, a relevant energy
scale is  $m c_p^2$ where $c_p$ is the sound velocity computed at the
center of the cloud,
which fulfills $mc_p^2=n\partial_n(\mu)|_{n=n_p}$.
For pure 1D quasicondensates,
$mc_p^2=g_{1D}n_p$ where $n_p$ is the peak density.
In our data sets the linear densities can reach values which
are not small compared to $1/a_{3D}$ 
and we use the more general expression $mc_p^2=n_pg_{1D}/\sqrt{1+4n_p a_{3D}}$.
The evolution of the energy $mc_p^2$ for the data sets is shown
in Fig.~(\ref{fig.evolNat})(b). The
variation of $mc_p^2$ during time is as large as a factor 2.5.

\begin{figure}
  \centerline{
  \includegraphics[height=6cm]{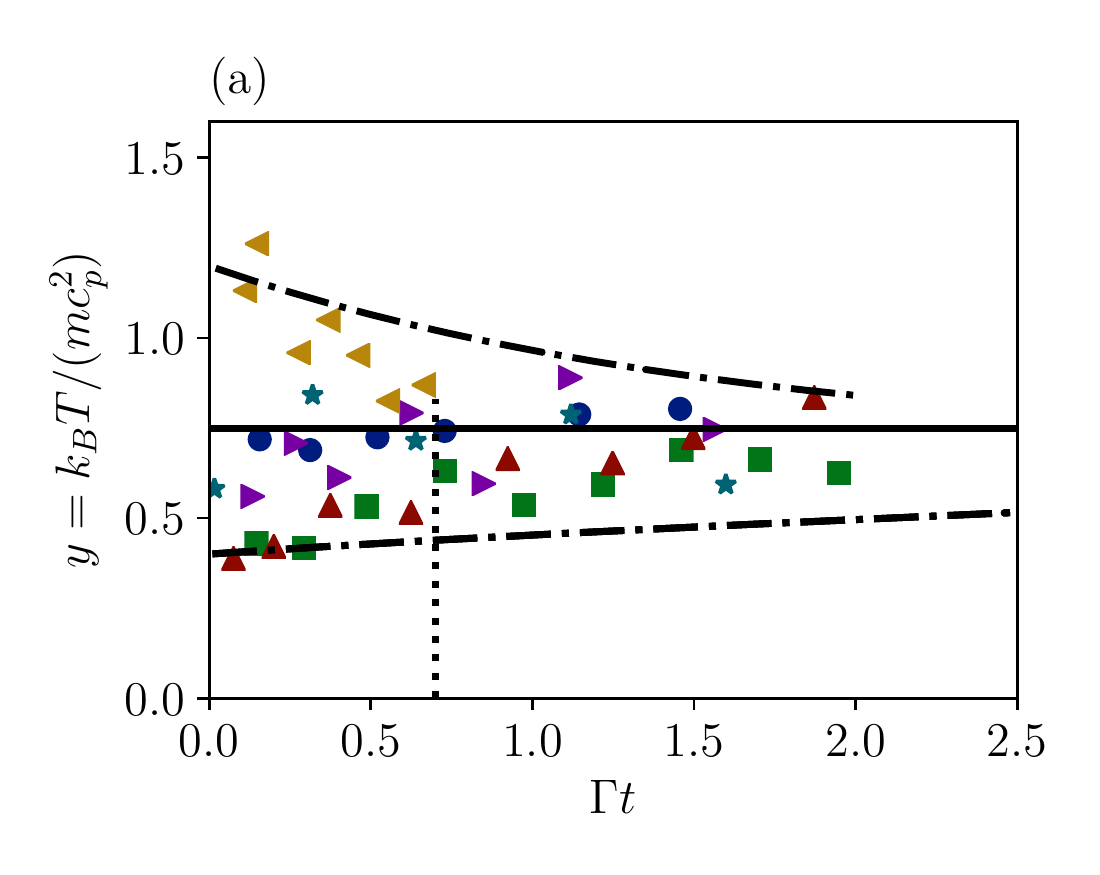}\includegraphics[height=6cm]{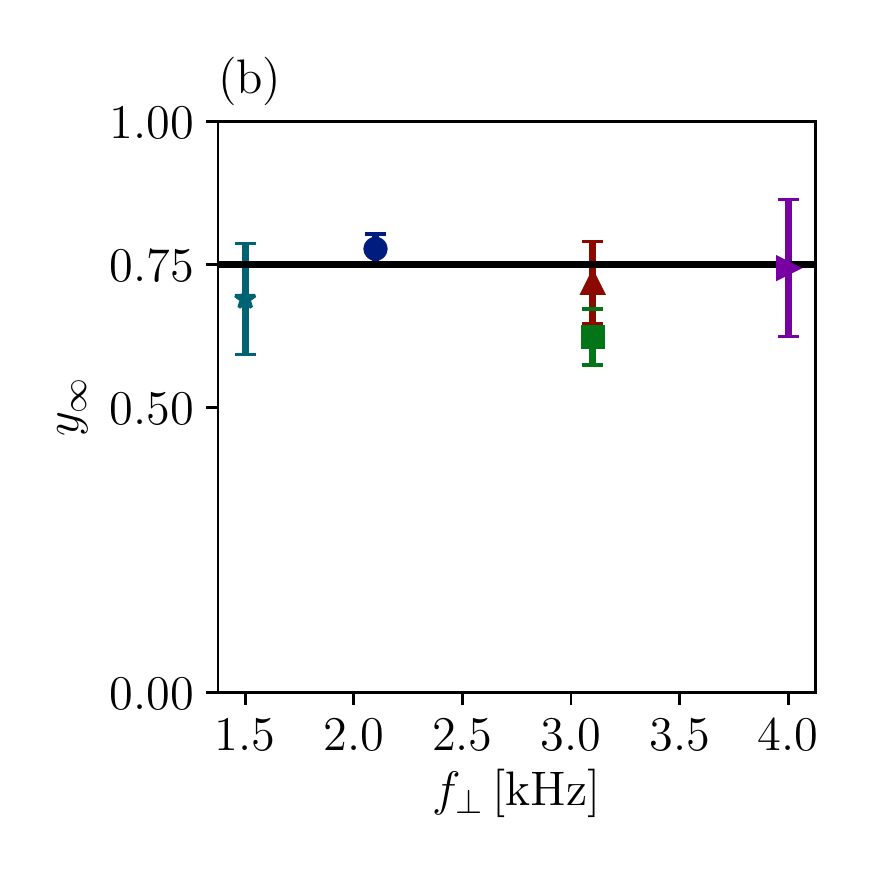}}
  \caption{
    (a) Time evolution of the ratio $y=k_B T/(mc_p^2)$ for the different data sets,
    shown as a function of $\Gamma t$. The solid horizontal line shows $y_\infty^{theo}=0.75$.
    The dashed-dotted lines are the computed expected time evolutions
     corresponding to initial situations of the data set 1 and 5.
    (b) For each data set, when available, mean value of $y$ for times
     larger than $0.7/\Gamma$. Error bars show the standard deviation among the data
  points that fulfill $\Gamma t > 0.7$.}
  \label{fig.evolTemp}
\end{figure}

The time evolution of the ratio $y=k_B T/(mc_p^2)$ is shown in Fig.~(\ref{fig.evolTemp})(a) for all
the data sets.
Theory for 1D harmonically confined gases~\cite{bouchoule_cooling_2018}
predicts that $y$ converge at long
times towards the asymptotic value $y_{\infty}^{\rm{theo}}=0.75$, shown as
solid black line in
Fig.~(\ref{fig.evolTemp})(a). 
The observed behavior is compatible with this prediction: the
spread of values of $y$ among different
data sets
decreases as $\Gamma t$ increases and 
at long times, all data gather around $y_{\infty}^{\rm{theo}}=0.75$, regardless of
the loss rate $\Gamma$ and of the transverse oscillation frequency.
For the data sets 2,3 and 6, $y$ deviates by no more
than 20\% from $y_\infty^{theo}$ over the
whole time evolution while
$mc_p^2$ decreases by a factor up to 2.5.
 For all data sets $y$ is about stationnary for times
  $t>0.7/\Gamma$ and we note
  $y_\infty$ the mean value of $y$ for times $t>0.7/\Gamma$.
Fig.~(\ref{fig.evolTemp})(c) shows $y_\infty$, plotted versus
the transverse oscillation frequency.
 Results are close to
$y_\infty^{theo}$, with $|y_\infty - y_\infty^{theo}|/y_\infty^{theo} < 0.2$.
The observed discrepancy between $y_\infty$ and $y_\infty^{theo}$
may be due on the one hand to our finite thermometry precision (the value $y_\infty$
is within one error bar of
$y_\infty$ for most data sets) and on the other hand to the fact that the criteria
$\Gamma t > 0.7$ might be insufficient to unsure
the asymptotic value of $y$ has be attained.

Quantitative experimental investigation of
the time-evolution of $y$ under the effect of losses is difficult with  our data sets.
The reason is  that the initial condition we produce are such
that the maximal deviation between $y(t=0)$ and $y_\infty^{theo}$ is comparable to
our thermometry resolution. 
We attribute this to the preparation scheme where, for our experimental procedure,  three-body
losses during the evaporative cooling probably impose a value of
$y$ close to 0.75~\cite{schemmer_cooling_2018}. 
On the theoretical side, for a given initial condition,
the expected time evolution of $y$ can be computed using the dynamical
equations derived in~\cite{bouchoule_cooling_2018}.
For pure 1D harmonically confined cloud the equation
reduces to $dy/d(\Gamma t)= y/3 + 1/4 $. To take into account
the 3D effect due to transverse swelling of the wave-function, we solved numerically the
general equations given in~\cite{bouchoule_cooling_2018}. 
We show in dashed-dotted black lines in Fig.~(\ref{fig.evolTemp})(a) the expected time-evolution
for initial conditions corresponding to the $1^{\rm st}$ and the $5^{\rm th}$ data sets.
The expected convergence of $y$ towards $y_\infty^{theo}$ is found to be very slow for the
data set 5. Here transverse swelling effects slow down the dynamics\footnote{Because of transverse
  swelling effect at large density, for some initial parameters, the function $y(t)$ could even be not
  monotonous. }.
Experimentally,
the convergence appears to be slightly faster.
For initial situations corresponding to the data set 1  on the other hand, transverse
swelling effects are expected to speed up the dynamics.
Data are consistent with this behavior.

\section{Conclusion}
In this paper, we identify for the first time the asymptotic temperature of a 1D quasicondensate
submitted to a 1-body loss process: more precisely, we show that the ratio $k_B T/(mc_p^2)$ reaches
an asymptotic value, close to the theoretical prediction of 0.75.
In a previous work~\cite{rauer_cooling_2016} which investigates the evolution of
the temperature of a quasicondensate
under the effect of losses, 1D quasicondensates were shown to reach
lower ratios $k_B T /(mc_p^2)$, in disagreement with theoretical predictions.
The difference between the work~\cite{rauer_cooling_2016} and the present work is two-fold. First, in
~\cite{rauer_cooling_2016} the out-coupling is realised with a monochromatic field, in which case,
for chemical potential of the order of the transverse trapping frequency, homogeneity of the loss
process is not guaranteed. Such an inhomogeneity makes the loss process sensitive to the energy of the atoms; a phenomena not accounted for by the model.
In this paper the use of a wide MW power spectrum ensures the homogeneity of the losses.  
 Second,
the data sets in~\cite{rauer_cooling_2016} for which ratios $y$ lower than expected are
reported, correspond to losses engineered via a radio-frequency field that couples magnetic states
within the hyperfine level $F=2$: in opposition to what happens when
using microwave outcoupling to $F=1$, 
the transfer of the trapped  atoms, which are in the 
$|F=2,m_F=2\rangle$ state, to untrapped states $|F=2,m_F'\leq 0 \rangle$
involves the intermediate state
$|F=2,m_F=1\rangle$ which is  held in the magnetic trap. Since both states contribute to the images and a priori host uncorrelated fluctuations, one expects a decrease of the density ripple power spectrum and thus of the fitted temperature.

This work leads to many open-questions.
First, 
the thermometry we use probes the collective modes which lie
in the phononic regime\footnote{Their  frequency
is much smaller than $\mu/\hbar$.}, while 
theoretical predictions~\cite{johnson_long-lived_2017}  indicate that collective modes
of higher frequency reach, under the effect of losses, higher temperatures. 
As already pointed out in~\cite{johnson_long-lived_2017}, for clouds confined
in a smoothly varying potential, information
on higher frequency collective modes may be retrieved from the wings of the cloud, namely the part
of the density profile that extends beyond the size of $R$ of a quasicondensate.
Indeed, as losses occur, we observe the growing of the fraction of atoms present in the
wings  and the density in the wings
typically largely exceed that  expected
for a cloud at thermal equilibrium at a temperature equal to the temperature extracted from 
the density ripple thermometry, 
as is shown in Fig.~(\ref{fig.drawing_analysis}).
This call for further theoretical and experimental investigations. 
Second, the theoretical prediction that for phonons the ratio $k_B T/(mc^2)$ reaches an
asymptotic value is {\it a priori} also valid in
higher dimensions.
It is an open question
whether coupling to higher frequency collective modes, an  inefficient
process in 1D\cite{johnson_long-lived_2017}, prevents the phonon modes
to attain this asymptotic behavior.
Finally, it would be interesting to extend the investigation of the effect of losses
to regimes different from the (quasi-)condensate regime. In the case
of 1D Bose gases with contact interactions, that are described by the Lieb-Liniger model,
a description in terms of the evolution of the distribution of rapidities~\cite{lieb_exact_1963}
would permit to generalize the studies to all possible states of the gas. 

\section{Acknowledgments}
The authors thanks Bernhard Rauer for interesting discussions and Marc Cheneau for his suggestion of
using micro-wave field to induce losses. 
M. S. gratefully acknowledges support by the Studienstiftung des deutschen Volkes.
This work was supported by Région Île de France (DIM NanoK, Atocirc project). The authors thank
Sophie Bouchoule of C2N (centre nanosciences 
et nanotechnologies, CNRS / UPSUD, Marcoussis, France) for the development 
and microfabrication of the atom chip. Alan Durnez and Abdelmounaim Harouri 
of C2N are acknowledged for their technical support. C2N laboratory is a 
member of RENATECH, the French national network of large facilities for 
micronanotechnology.


\bibliography{papier1corps.bib}

\begin{thebibliography}{10}
\providecommand{\url}[1]{\texttt{#1}}
\providecommand{\urlprefix}{URL }
\expandafter\ifx\csname urlstyle\endcsname\relax
  \providecommand{\doi}[1]{doi:\discretionary{}{}{}#1}\else
  \providecommand{\doi}{doi:\discretionary{}{}{}\begingroup
  \urlstyle{rm}\Url}\fi
\providecommand{\eprint}[2][]{\url{#2}}

\bibitem{barreiro_open-system_2011}
J.~T. Barreiro, M.~Müller, P.~Schindler, D.~Nigg, T.~Monz, M.~Chwalla,
  M.~Hennrich, C.~F. Roos, P.~Zoller and R.~Blatt,
\newblock \emph{An open-system quantum simulator with trapped ions},
\newblock Nature \textbf{470}(7335), 486 (2011),
\newblock \doi{10.1038/nature09801}.

\bibitem{tomita_observation_2017}
T.~Tomita, S.~Nakajima, I.~Danshita, Y.~Takasu and Y.~Takahashi,
\newblock \emph{Observation of the {Mott} insulator to superfluid crossover of
  a driven-dissipative {Bose}-{Hubbard} system},
\newblock Science Advances \textbf{3}(12), e1701513 (2017),
\newblock \doi{10.1126/sciadv.1701513}.

\bibitem{barontini_controlling_2013}
G.~Barontini, R.~Labouvie, F.~Stubenrauch, A.~Vogler, V.~Guarrera and H.~Ott,
\newblock \emph{Controlling the {Dynamics} of an {Open} {Many}-{Body} {Quantum}
  {System} with {Localized} {Dissipation}},
\newblock Physical Review Letters \textbf{110}(3), 035302 (2013),
\newblock \doi{10.1103/PhysRevLett.110.035302}.

\bibitem{wouters_excitations_2007}
M.~Wouters and I.~Carusotto,
\newblock \emph{Excitations in a {Nonequilibrium} {Bose}-{Einstein}
  {Condensate} of {Exciton} {Polaritons}},
\newblock Physical Review Letters \textbf{99}(14), 140402 (2007),
\newblock \doi{10.1103/PhysRevLett.99.140402}.

\bibitem{wouters_spatial_2008}
M.~Wouters, I.~Carusotto and C.~Ciuti,
\newblock \emph{Spatial and spectral shape of inhomogeneous nonequilibrium
  exciton-polariton condensates},
\newblock Physical Review B \textbf{77}(11), 115340 (2008),
\newblock \doi{10.1103/PhysRevB.77.115340}.

\bibitem{keeling_spontaneous_2008}
J.~Keeling and N.~G. Berloff,
\newblock \emph{Spontaneous {Rotating} {Vortex} {Lattices} in a {Pumped}
  {Decaying} {Condensate}},
\newblock Physical Review Letters \textbf{100}(25), 250401 (2008),
\newblock \doi{10.1103/PhysRevLett.100.250401}.

\bibitem{lagoudakis_quantized_2008}
K.~G. Lagoudakis, M.~Wouters, M.~Richard, A.~Baas, I.~Carusotto, R.~André,
  L.~S. Dang and B.~Deveaud-Plédran,
\newblock \emph{Quantized vortices in an exciton–polariton condensate},
\newblock Nature Physics \textbf{4}(9), 706 (2008),
\newblock \doi{10.1038/nphys1051}.

\bibitem{carusotto_spontaneous_2005}
I.~Carusotto and C.~Ciuti,
\newblock \emph{Spontaneous microcavity-polariton coherence across the
  parametric threshold: {Quantum} {Monte} {Carlo} studies},
\newblock Physical Review B \textbf{72}(12), 125335 (2005),
\newblock \doi{10.1103/PhysRevB.72.125335}.

\bibitem{wouters_stochastic_2009}
M.~Wouters and V.~Savona,
\newblock \emph{Stochastic classical field model for polariton condensates},
\newblock Physical Review B \textbf{79}(16), 165302 (2009),
\newblock \doi{10.1103/PhysRevB.79.165302}.

\bibitem{grisins_degenerate_2016}
P.~Grišins, B.~Rauer, T.~Langen, J.~Schmiedmayer and I.~E. Mazets,
\newblock \emph{Degenerate {Bose} gases with uniform loss},
\newblock Physical Review A \textbf{93}(3), 033634 (2016),
\newblock \doi{10.1103/PhysRevA.93.033634}.

\bibitem{bouchoule_cooling_2018}
I.~Bouchoule, M.~Schemmer and C.~Henkel,
\newblock \emph{Cooling phonon modes of a {Bose} condensate with uniform few
  body losses},
\newblock SciPost Physics \textbf{5}(5), 043 (2018),
\newblock \doi{10.21468/SciPostPhys.5.5.043}.

\bibitem{szymanska_nonequilibrium_2006}
M.~H. Szymańska, J.~Keeling and P.~B. Littlewood,
\newblock \emph{Nonequilibrium {Quantum} {Condensation} in an {Incoherently}
  {Pumped} {Dissipative} {System}},
\newblock Physical Review Letters \textbf{96}(23), 230602 (2006),
\newblock \doi{10.1103/PhysRevLett.96.230602}.

\bibitem{schemmer_cooling_2018}
M.~Schemmer and I.~Bouchoule,
\newblock \emph{Cooling a {Bose} {Gas} by {Three}-{Body} {Losses}},
\newblock Physical Review Letters \textbf{121}(20), 200401 (2018),
\newblock \doi{10.1103/PhysRevLett.121.200401}.

\bibitem{rauer_cooling_2016}
B.~Rauer, P.~Grišins, I.~Mazets, T.~Schweigler, W.~Rohringer, R.~Geiger,
  T.~Langen and J.~Schmiedmayer,
\newblock \emph{Cooling of a {One}-{Dimensional} {Bose} {Gas}},
\newblock Physical Review Letters \textbf{116}(3), 030402 (2016),
\newblock \doi{10.1103/PhysRevLett.116.030402}.

\bibitem{schemmer_out--equilibrium_2019}
M.~Schemmer,
\newblock \emph{Out-of-equilibrium dynamics in {1D} {Bose} gases}  (2019).

\bibitem{trebbia_roughness_2007}
J.-B. Trebbia, C.~L. Garrido~Alzar, R.~Cornelussen, C.~I. Westbrook and
  I.~Bouchoule,
\newblock \emph{Roughness {Suppression} via {Rapid} {Current} {Modulation} on
  an {Atom} {Chip}},
\newblock Physical Review Letters \textbf{98}(26), 263201 (2007),
\newblock \doi{10.1103/PhysRevLett.98.263201}.

\bibitem{kheruntsyan_pair_2003}
K.~V. Kheruntsyan, D.~M. Gangardt, P.~D. Drummond and G.~V. Shlyapnikov,
\newblock \emph{Pair {Correlations} in a {Finite}-{Temperature} 1d {Bose}
  {Gas}},
\newblock Physical Review Letters \textbf{91}(4), 040403 (2003),
\newblock \doi{10.1103/PhysRevLett.91.040403}.

\bibitem{mora_extension_2003}
C.~Mora and Y.~Castin,
\newblock \emph{Extension of {Bogoliubov} theory to quasicondensates},
\newblock Physical Review A \textbf{67}(5), 053615 (2003),
\newblock \doi{10.1103/PhysRevA.67.053615}.

\bibitem{salasnich_effective_2002}
L.~Salasnich, A.~Parola and L.~Reatto,
\newblock \emph{Effective wave equations for the dynamics of cigar-shaped and
  disk-shaped {Bose} condensates},
\newblock Physical Review A \textbf{65}(4), 043614 (2002),
\newblock \doi{10.1103/PhysRevA.65.043614}.

\bibitem{fuchs_hydrodynamic_2003}
J.~N. Fuchs, X.~Leyronas and R.~Combescot,
\newblock \emph{Hydrodynamic modes of a one-dimensional trapped {Bose} gas},
\newblock Physical Review A \textbf{68}(4), 043610 (2003),
\newblock \doi{10.1103/PhysRevA.68.043610}.

\bibitem{olshanii_atomic_1998}
M.~Olshanii,
\newblock \emph{Atomic {Scattering} in the {Presence} of an {External}
  {Confinement} and a {Gas} of {Impenetrable} {Bosons}},
\newblock Physical Review Letters \textbf{81}(5), 938 (1998),
\newblock \doi{10.1103/PhysRevLett.81.938}.

\bibitem{dettmer_observation_2001}
S.~Dettmer, D.~Hellweg, P.~Ryytty, J.~J. Arlt, W.~Ertmer, K.~Sengstock, D.~S.
  Petrov, G.~V. Shlyapnikov, H.~Kreutzmann, L.~Santos and M.~Lewenstein,
\newblock \emph{Observation of {Phase} {Fluctuations} in {Elongated}
  {Bose}-{Einstein} {Condensates}},
\newblock Physical Review Letters \textbf{87}(16), 160406 (2001),
\newblock \doi{10.1103/PhysRevLett.87.160406}.

\bibitem{imambekov_density_2009}
A.~Imambekov, I.~E. Mazets, D.~S. Petrov, V.~Gritsev, S.~Manz, S.~Hofferberth,
  T.~Schumm, E.~Demler and J.~Schmiedmayer,
\newblock \emph{Density ripples in expanding low-dimensional gases as a probe
  of correlations},
\newblock Physical Review A \textbf{80}(3), 033604 (2009),
\newblock \doi{10.1103/PhysRevA.80.033604}.

\bibitem{manz_two-point_2010}
S.~Manz, R.~Bücker, T.~Betz, C.~Koller, S.~Hofferberth, I.~E. Mazets,
  A.~Imambekov, E.~Demler, A.~Perrin, J.~Schmiedmayer and T.~Schumm,
\newblock \emph{Two-point density correlations of quasicondensates in free
  expansion},
\newblock Physical Review A \textbf{81}(3), 031610 (2010),
\newblock \doi{10.1103/PhysRevA.81.031610}.

\bibitem{schemmer_monitoring_2018}
M.~Schemmer, A.~Johnson and I.~Bouchoule,
\newblock \emph{Monitoring squeezed collective modes of a one-dimensional
  {Bose} gas after an interaction quench using density-ripple analysis},
\newblock Physical Review A \textbf{98}(4), 043604 (2018),
\newblock \doi{10.1103/PhysRevA.98.043604}.

\bibitem{johnson_long-lived_2017}
A.~Johnson, S.~S. Szigeti, M.~Schemmer and I.~Bouchoule,
\newblock \emph{Long-lived nonthermal states realized by atom losses in
  one-dimensional quasicondensates},
\newblock Physical Review A \textbf{96}(1), 013623 (2017),
\newblock \doi{10.1103/PhysRevA.96.013623}.

\bibitem{lieb_exact_1963}
E.~H. Lieb and W.~Liniger,
\newblock \emph{Exact {Analysis} of an {Interacting} {Bose} {Gas}. {I}. {The}
  {General} {Solution} and the {Ground} {State}},
\newblock Physical Review \textbf{130}(4), 1605 (1963),
\newblock \doi{10.1103/PhysRev.130.1605}.

\end{thebibliography}

\end{document}